\documentclass[12pt]{article}

\usepackage{amsmath}
\usepackage{amsfonts}
\usepackage{amssymb}
\usepackage{graphics,graphicx,tikz}
\usepackage[linktocpage]{hyperref}

\numberwithin{equation}{section} 
\def\beq{\begin{equation}}
\def\eeq{\end{equation}}
\def\Ord{{\cal O}}
\def\eqn#1{Eq.~(\ref{#1})}
\def\bea{\begin{align}}
\def\eea{\end{align}}

\usepackage{color}

\begin{document}
\begin{titlepage}
\hfill \hbox{CERN-TH-2020-143}
\vskip 0.1cm
\hfill \hbox{NORDITA 2020-079}
\vskip 0.1cm
\hfill \hbox{QMUL-PH-20-25}
\vskip 0.5cm
\begin{flushright}
\end{flushright}
\begin{center}
{\Large \bf Universality of ultra-relativistic \\ gravitational scattering} 
\vskip 1.0cm {\large  Paolo Di Vecchia$^{a, b}$, Carlo Heissenberg$^{b,c}$,
Rodolfo Russo$^{d}$, \\ Gabriele Veneziano$^{e, f}$ } \\[0.7cm]

{\it $^a$ The Niels Bohr Institute, University of Copenhagen, Blegdamsvej 17, \\DK-2100 Copenhagen, Denmark}\\
{\it $^b$ NORDITA, KTH Royal Institute of Technology and Stockholm University, \\
 Roslagstullsbacken 23, SE-10691 Stockholm, Sweden}\\
{\it $^c$ Department of Physics and Astronomy, Uppsala University,\\ Box 516, SE-75120 Uppsala, Sweden}\\
{\it $^d$ Centre for Research in String Theory, School of Physics and Astronomy \\ Queen Mary University of London, Mile End Road, E1 4NS London, United Kingdom}\\
{\it $^e$ Theory Department, CERN, CH-1211 Geneva 23, Switzerland}\\
{\it $^f$Coll\`ege de France, 11 place M. Berthelot, 75005 Paris, France}
\end{center}
\begin{abstract}
  We discuss the ultra-relativistic gravitational scattering of two massive particles at two-loop (3PM) level. We find that in this limit the real part of the eikonal, determining the  deflection angle, is universal for gravitational theories in the two derivative approximation. This means that, regardless of the number of supersymmetries or the nature of the probes, the result connects smoothly with the massless case discussed since the late eighties by Amati, Ciafaloni and Veneziano. We analyse the problem both by using the analyticity and crossing properties of the scattering amplitudes and, in the case of the maximally supersymmetric theory, by explicit evaluation of the 4-point 2-loop amplitude using the results for the integrals in the full soft region. The first approach shows that the observable we are interested in is determined by the inelastic tree-level amplitude describing the emission of a graviton in the high-energy double-Regge limit, which is the origin of the universality property mentioned above. The second approach strongly suggests that the inclusion of the whole soft region is a necessary (and possibly sufficient) ingredient for recovering ultra relativistic finiteness and universality at the 3PM level. We conjecture that this universality persists at all orders in the PM expansion.
\end{abstract}
\end{titlepage}


\section{Introduction}

At high energy and large impact parameter, gravitational scattering is dominated by the exchange of the highest-spin massless states in the theory~\cite{tHooft:1987vrq, Amati:1987wq,Amati:1987uf,Muzinich:1987in,Sundborg:1988tb}. The focus of this work is on standard gravitational field theories in the two derivative approximation, so the highest spin  particle is the graviton. The observable we study is the elastic $2 \to 2$ amplitude in the ultra-relativistic regime, where the centre-of-mass energy $E_\textrm{cm}=\sqrt{s}$ is much larger than any other energy scale in the process (which goes under the name of Regge limit). Notice that this implies that the impact parameter $b$, which is related by Fourier transform to the total momentum exchanged in the scattering, is much larger than the gravitational length scale\footnote{As usual, working in a $D$-dimensional theory is also convenient  for regularising the IR divergences of the $D=4$ case.} $R^{D-3} \sim G_N\sqrt{s}$, defined in analogy with the Schwarzschild radius. Since the process is dominated by the exchange of gravitons, it is natural to expect a universal result for the high energy gravitational scattering. It is well known that this is the case to first order in $R/b\ll 1$ where the result is captured by the leading eikonal phase $\delta_0$ irrespectively of whether one works with strings or point particles, in higher dimensions, or in the presence of supersymmetry ~\cite{Amati:1987wq,Amati:1987uf}. It is then natural to ask whether such universality persists at higher orders in $R/b$, at least in the class of theories mentioned above\footnote{Note that, while at this order in $R/b$ the problem can be studied, following~\cite{tHooft:1987vrq}, as the scattering of one particle in the Aichelburg-Sexl 
shock-wave produced by the other, such a simple picture fails at higher orders.}. 

The first non-trivial contribution to the ultra-relativistic result appears at sub-sub-leading order in the small $R/b$ limit which captures the 3PM (post-Minkowskian) correction. This contribution can be encoded in terms of a correction $\delta_2$ to the full eikonal. A first novelty with respect to the leading result is that $\delta_2$ has both a real and an imaginary part, so ${\rm e}^{2i \delta}$ is not just a phase. The imaginary part is related to non-conservative processes, such as the bremsstrahlung emission of massless particles, while the real part captures the conservative dynamics. A first result for $\delta_2$ was obtained in~\cite{Amati:1990xe} for the scattering in pure general relativity (GR) of two massless scalars representing, classically, the collision of {\it two}  Aichelburg-Sexl 
shock-waves. The analysis of~\cite{Bellini:1992eb} suggested that the same result for $\operatorname{Re}(\delta_2)$ should hold also for supergravity theories and this was explicitly verified in~\cite{DiVecchia:2019kta} for the case of the maximally supersymmetric theory. Explicit comparison between the GR and the ${\cal N}=8$ results shows that $\operatorname{Im}(\delta_2)$ is not universal (although a crucial $\log(s)$-enhanced term is), which is expected since the massless spectrum of the two theories is different. The universality of the 3PM conservative dynamics in the massless case has been confirmed and extended~\cite{Bern:2020gjj} to theories with different amounts of supersymmetry. 

A similar, but slightly different setup is to consider the scattering of two scalar particles of masses $m_1$ and $m_2$. There has been considerable interest, recently, in the use of new scattering amplitude techniques for extracting information about the conservative scattering process in the case $s\sim m_i^2$~\cite{Cheung:2018wkq,Bjerrum-Bohr:2018xdl,Bern:2019nnu,KoemansCollado:2019ggb,Bern:2019crd,Bjerrum-Bohr:2019kec,Kalin:2019rwq,Kalin:2019inp,Cristofoli:2020uzm}: the aim is to pin down the relevant ingredients needed for describing the inspiral phase of black-hole mergers. The importance of such a result for computing the templates for actual gravitational-wave experiments have been stressed, in particular by Damour~\cite{Damour:2017zjx}.  

The first result at 3PM level was derived a little while ago in \cite{Bern:2019nnu,Bern:2019crd} and was  later confirmed in \cite{Cheung:2020gyp,Kalin:2020fhe}, while the case of maximally supersymmetric gravity was obtained recently in~\cite{Parra-Martinez:2020dzs}. The ultra-relativistic limit of these massive results is qualitatively different from the massless case mentioned above, as the leading term of $\operatorname{Re}(\delta_2)$ has a $\log (s/(m_1 m_2))$ enhancement with respect to the one obtained in~\cite{Amati:1990xe}. 
 From the amplitude perspective the origin of this $\log (s/(m_1 m_2))$-enhanced contribution lies in a particular scalar integral which appears in exactly the same way in both the ${\cal N}=0$ and the ${\cal N}=8$ cases, hence the proposal of~\cite{Parra-Martinez:2020dzs} that there is a separate universality class for the logarithmically-enhanced term of the massive scattering which is different from the massless case where there is no $\log(s)$-enhancement.

This unexpected result of  \cite{Bern:2019nnu,Bern:2019crd} has provoked quite a lot of discussion since, taken at face value, it would lead to a divergent deflection angle in either the massless or the ultra-relativistic  limit. Since gravity is known \cite{Weinberg:1965nx} to be free of mass/collinear divergences the authors of  \cite{Bern:2019nnu,Bern:2019crd}
 have immediately pointed out that their result only holds for sufficiently small values of the ratio $\frac{q}{m}$ where $q \sim \frac{\hbar}{b}$ is the momentum transfer in the perturbative amplitude.
 They have also given  \cite{Bern:2019crd} a one-loop example (involving a non-classical term) of how the $m \to 0$ and the $q \to 0$ limits can be quite different.
 On the other hand the above-mentioned divergence is also present at finite $m$ for $s \to \infty$ and would persist even if, for $m > q$, one would replace the $\log (s/(m_1 m_2))$ by a $\log (s/q^2)$. There is also some tension with  expectations based on the ``self-force'' approach to PM dynamics. On these different grounds Damour suggested that $\log (s/(m_1 m_2))$ enhancement $\operatorname{Re}(\delta_2)$ cannot be present in the ultra-relativistic limit. He proposed in~\cite{Damour:2019lcq} a modification of the result of \cite{Bern:2019nnu,Bern:2019crd}  with a smooth massless limit that however differs (even in sign!) from the one of~\cite{Amati:1990xe}.  Finally, a check~\cite{Blumlein:2020znm, Bini:2020wpo} proposed in \cite{Damour:2019lcq} for distinguishing the two alternatives at  6PN order has contradicted his original proposal while it is consistent with  \cite{Bern:2019nnu,Bern:2019crd}. Other alternatives, still at variance with \cite{Amati:1990xe}, have been proposed in a later version of \cite{Damour:2019lcq}.

As mentioned above, in this short note we will reassess the ultra-relativistic limit $s\gg m^2_i$ and ask whether one recovers the massless shock-wave result for the real part of the eikonal $\delta$ at 3PM level.
We will address this issue by using two complementary techniques. The first approach follows the argument of~\cite{Amati:1990xe} where the 3PM result was derived by exploiting the analyticity and crossing properties of the scattering amplitudes among scalars. These properties imply a dispersion relation connecting the leading energy behaviour of the imaginary and real parts of the $2\to 2$ amplitude. Then, in this limit, the conservative 3PM dynamics can be derived from a particular unitarity cut of the two-loop amplitude, thus replacing the full-fledged 2-loop calculation by a phase-space integral of a product of two tree-level amplitudes. As an aside, notice that this provides an explicit check that the 3PM conservative dynamics is entirely determined by classical on-shell data. The second approach follows the direct loop amplitude derivation of~\cite{Bern:2019nnu,Bern:2019crd,Parra-Martinez:2020dzs}, where we just reconsider the calculation of the integrals without relying on the ``potential'' approximation.
We use the eikonal approach to extract the classical contribution, instead of the effective field theory comparison of~\cite{Cheung:2018wkq}, and we work with the full result of the integrals in the ``soft region'' (i.e. in the limit of small momentum transfer). Both the approach based on analyticity/crossing and the one using the explicit loop amplitudes yield the same $\operatorname{Re}(\delta_2)$ which also agrees with the massless result of~\cite{Amati:1990xe,DiVecchia:2019kta,Bern:2020gjj}, thus suggesting that the origin of the different ultra-relativistic behaviour of~\cite{Bern:2019nnu,Bern:2019crd,Cheung:2020gyp,Parra-Martinez:2020dzs,Kalin:2020fhe} lies in the use of the ``potential'' approximation in evaluating the loop integrals. When including the full contribution of the soft region, the ultra-relativistic limit at 3PM order is universal regardless of the mass of the external states. We conjecture that this is actually true to all orders in the PM expansion, since the relevant contributions are described by diagrams where the external states emit only gravitons and the two highly boosted lines representing them are connected through a tree-level GR amplitude~\cite{Amati:2007ak}. Both ingredients are universal in the class of theories we focus on.

The paper is organized as follows. In Section~\ref{analyticity} we revisit the analyticity/crossing of~\cite{Amati:1990xe} deriving on general grounds a relation between the ultra-relativistic limit of $\operatorname{Re}(\delta_2)$ and the leading high energy behaviour of the imaginary part of the 2-loop amplitude. The analysis holds for arbitrary values of the masses as long as $s\gg m_i^2$ and also shows that the ultra-relativistic limit of $\operatorname{Re}(\delta_2)$ can be universal only if $\operatorname{Im}(\delta_2)$ is enhanced by a single power of $\log(s)$.
In Section~\ref{highenergy}, we again follow~\cite{Amati:1990xe} and evaluate the $2\to 3$ tree-level amplitude relevant for the 3-particle cut determining the imaginary part of the 2-loop amplitude. Since we are interested in the ultra-relativistic case, we focus on a double Regge limit of the tree-level amplitude and show that, in this regime, it is universal and equal to the massless result of~\cite{Amati:1990xe,Ademollo:1990sd}. This implies that the ultra-relativistic limit of $\operatorname{Im}(\delta_2)$ is enhanced by a single factor of $\log(s)$ and that the ultra-relativistic limit of $\operatorname{Re}(\delta_2)$ is also universal and has no enhancement at all. In Section~\ref{direct} we follow the approach of~\cite{Bern:2019nnu,Bern:2019crd} and in particular focus on the massive ${\cal N}=8$ case studied in~\cite{Parra-Martinez:2020dzs}. We provide the results for the integrals necessary to calculate the 3PM massive eikonal in ${\cal N}=8$ supergravity in the ultra-relativistic limit. As mentioned, we consider the full ``soft'' region result and show that the universal ultra-relativistic result for $\operatorname{Re}(\delta_2)$ of Section~\ref{highenergy} is recovered. We also provide the result for $\operatorname{Re}(\delta_2)$ and the 3PM deflection angle for the generic values of $s/m_i^2$ so as to facilitate the comparison with~\cite{Parra-Martinez:2020dzs} even if we leave the discussion of the derivation to another work \cite{toap}. In Appendix~\ref{Appa} we provide some further details on the consequences of analyticity and crossing used  in the main text. In Appendix~\ref{Appb} we collect all the results we need for the high-energy scalar integrals in the soft region.  

\section{The analyticity/crossing argument revisited}
\label{analyticity}

In this section we review, improve and extend the arguments given in \cite{Amati:1990xe} for connecting  the leading high-energy expression of $\operatorname{Re}(\delta_2)$  to the inelastic contribution to the imaginary part of the two loop scattering amplitude. The latter is the convolution of two on-shell tree amplitudes and thus an easier object to deal with.
For definiteness we will  consider the case of the elastic gravitational scattering of two non-identical scalar particles of mass $m_1, m_2$. Crossing symmetry under the exchange of the two Mandelstam variables $s$ and $u$ simplifies considerably our analysis although we believe that the final results generalize to elastic processes that lack exact $s \leftrightarrow u$ symmetry.

The essential ingredients of the argument presented in~\cite{Amati:1990xe} are:
\begin{enumerate}
\item[i)] real-analyticity of the scattering amplitude $A(s^*, t) = A^*(s, t)$ as a function of the complex variable $s$ at $t \le 0$, where $-t=q^2$ is the exchanged momentum  (squared); 
\item[ii)] its $s \leftrightarrow u $ crossing symmetry $A(s,t) = A(u, t)$ with $u = - s -t + 2 (m_1^2 + m_2^2)$; 
\item[iii)] some information about its high-energy asymptotic behaviour at each loop order, a property that fixes the number of subtractions needed in order to write a convergent dispersion relation for $A$, see Appendix A. 
\end{enumerate}We will also make use of the exponentiation in impact-parameter space  needed for recovering a classical limit.

We will be helped by an amusing mathematical analogy with high-energy hadronic (QCD) scattering whereby the elastic hadronic amplitude $A^{\rm Had}(s,t)$ is believed to behave, at high energy, as $s \log^p s$ with some (not necessarily integer) $p$. An important quantity discussed in that context  \cite{Bronzan:1974jh}, \cite{Kang:1974gt} (see \cite{Fagundes:2017iwb} for a recent review) is  the ratio $\frac{\operatorname{Re} A^{\rm Had}(s,0)}{\operatorname{Im} A^{\rm Had}(s,0)}$. Such a ratio (extended to the case of non-vanishing $t$) plays an important role in \cite{Amati:1990xe} as we shall see hereafter. Note that constraints based on analyticity and crossing, being linear, apply at each loop order, at each order in $\epsilon$, and also separately to different terms in the high-energy expansion (like in the hadronic case where the Pomeron may be accompanied by subleading Regge pole contributions).
Since it is not easy to find a self-contained account of this methodology in the literature, and we are not constrained here by the Froissart bound, we will sketch for completeness the basic argument in Appendix A and refer to the above literature for further details.

We start by recalling the expression for the high energy tree-level amplitude $A_0(s,t)$ and the corresponding eikonal phase  $2 \delta_0$ for generic $D = 4- 2 \epsilon$ keeping only the first relevant terms in an expansion in inverse powers of $s$. One finds \cite{Kabat:1992tb,KoemansCollado:2019ggb}:
\beq
A_0(s,t) = - \frac{8 \pi G_N s^2 \left(1- \frac{\Sigma}{s}+ \Ord(s^{-2})\right)}{t} +\text{analytic terms in }t \, ,
\label{A0}
\eeq
where $\Sigma \equiv 2\left(m_1^2 + m_2^2\right)$ and correspondingly\footnote{The kinematics is discussed in more detail before~\eqref{PHI10}. In our conventions the eikonal phase is $2 \delta$ and is given by the Fourier transform of $\frac{1}{4 E_\textrm{cm} P } A(s,t)$, with $2E_\textrm{cm} P = 2\sqrt{(p_1 p_2)^2-m_1^2 m_2^2}= s(1 - \frac{\Sigma}{2s} +  \Ord(s^{-2}))$ is the product of the centre-of-mass energy and momentum. Following \cite{KoemansCollado:2019ggb}, we denote by $\tilde{M}(s,b)$ the Fourier transform of $\frac{1}{4 E_\textrm{cm} P} M(s,t)$ for a generic function $M(s,t)$.}:
\beq
2 \delta_0 = \tilde{A}_0 \equiv \int \frac{d^{D-2} \vec{q}}{(2\pi)^{D-2}} \frac{A_0\left(s,t=-\vec{q}^{\;2}\right)\,e^{i\vec q \cdot \vec b}}{4\sqrt{(p_1 p_2)^2-m_1^2 m_2^2}} = G_N s \left(1- \frac{\Sigma}{2s}+ \Ord(s^{-2})\right) \Gamma(- \epsilon)(\pi b^2)^{\epsilon}\, .
\label{delta0}
\eeq
The high-energy behaviour of $\tilde{A}_0$ gives the asymptotic form of the leading eikonal phase:
\begin{equation}
  \label{eq:d0def}
  2 \delta_0 \simeq G_N s  \Gamma(- \epsilon)(\pi b^2)^{\epsilon}\, .
\end{equation}

In the Regge limit the full amplitude is encoded in the expression \cite{DiVecchia:2019kta}\footnote{Since we will focus on a specific $s-u$-symmetric amplitude here we follow slightly different conventions from Eq.~(2.15) ~\cite{DiVecchia:2019kta} where the tree-level structure ${\hat{A}}^{(0)}$ was factorised.}
\begin{equation}
  \label{eq:25}
  i\tilde{A} = \left(1+2 i \Delta(s,b) \right) {\rm e}^{2 i \delta} - 1\;,
\end{equation}
where $\delta=\delta_0 +\delta_1+\delta_2+ \cdots$ is the classical eikonal  and $\Delta$ encodes the quantum corrections. At one-loop  we know that $A$ must include a leading imaginary contribution, growing like $s^3$ (without extra logs) and responsible for the start of the exponentiation of $2 \delta_0$.  This indeed comes out of the box plus crossed box contribution in the form  \cite{KoemansCollado:2019ggb}:
\beq
\operatorname{Im} A_1^{(1)}(s,t) =  s^3 \left(1- \frac{3 \Sigma}{2s}+ \Ord\left(s^{-2}\right)\right)F_1(t, m_i^2)\,;~  {\rm with} ~ s^3 \left(1- \frac{3 \Sigma}{2s}\right) \tilde{F}_1 
= \frac{(2 \delta_0)^2 }{2}\, ,
\label{delta02}
\eeq
which therefore fixes $F_1$ modulo analytic terms as $t\to 0$. As explained in 
Appendix A analyticity and crossing symmetry imply 
 the following form for the amplitude itself:
\begin{align}
 & A_1^{(1)}(s,t) =
-\frac{1}{\pi} \left[ \left(s - \frac{\Sigma}{2}\right)^3  \log (-s) +  \left(u - \frac{\Sigma}{2}\right)^3  \log (-u) \right] F_1 (t,m_i^2) \nonumber \\
& \sim   s^3 \left(1 - \frac{3\Sigma}{2s}\right) F_1(t,m_i^2) \left(i + \frac{3 t}{\pi s} \left(\log s + \Ord(1)\right) \right)  +  \Ord(\Sigma^2 s,t^2 s)\, ,
\label{A11}
\end{align}
as  confirmed by explicit calculations \cite{KoemansCollado:2019ggb}.

It is known \cite{Amati:1990xe}  that, in order to compute the classical two loop contribution to the eikonal, one has to take into account the first quantum contribution at one-loop level up to $\Ord(\epsilon)$. It is also known from explicit calculations \cite{Henn:2019rgj, KoemansCollado:2019ggb,DiVecchia:2019myk, DiVecchia:2019kta} that such a contribution contains two powers of $s$ and  no $\log s$ in its  imaginary  part\footnote{There are also terms with an $s^{-\epsilon}$ behaviour in the  analytic part of the amplitude  \cite{Henn:2019rgj, DiVecchia:2019kta}. Although they are irrelevant for our subsequent discussion, we have checked that they also fulfil the constraints of analyticity and crossing.}. Using \eqn{neven} with $p=0$ from Appendix A we get:
\beq
A_1^{(2)}(s,t) 
 \sim    s^2 G_1(t, m_i^2) (-i \pi  + 2  \log s)  + \Ord(s^2) \, .
\label{A12}
\eeq
In \cite{Amati:1990xe} this second structure was omitted since, in the specific case discussed there, there was no correction to the leading imaginary part appearing in \eqn{A11} for $D=4$. This was also shown \cite{KoemansCollado:2019ggb} to be the case for non-vanishing\footnote{We have checked that the structure of  \eqn{A11} and \eqn{A12} is exactly recovered, at all $D$, by taking the high-energy limit of the explicit results (2.21), (2.25) given in \cite{KoemansCollado:2019ggb}.} $m_{1,2}$. However in other processes, in $D \ne 4$, or in supergravity theories, the structure \eqn{A12} is also present. We shall see that, provided a certain exponent takes a particular value, $A_1^{(2)}$ actually drops out  in the final expression for  $\operatorname{Re}(\delta_2)$, in agreement with several explicit calculations \cite{DiVecchia:2019kta,Bern:2020gjj}.

We finally turn to the two-loop amplitude $A_2$ . We know that it should have a leading $\Ord(s^4)$ real term in order to reproduce the correct third-order term in the exponentiation of $\delta_0$. Analyticity and crossing symmetry now fixes the amplitude to be of the form (see again Appendix A):
\begin{equation}
\begin{gathered}
 A_2^{(1)}(s,t) =  \left(s^4  + 2 s^3 (t - \Sigma)+ \Ord(t^2s^2, \Sigma^2 s^2)\right)  F_2(t, m_i^2) \\ 
 {\rm with}~ s^4\left(1- 2 \frac{\Sigma}{s}\right) \tilde{F}_2(b, m_i^2) = - \frac16 (2 \delta_0)^3 \, .
\end{gathered}
\label{A21}
\end{equation}
As a consequence $F_2(t,m_i^2)$ is known (up to analytic terms at $t=0$). Note that the terms $\Ord(s^2)$ in \eqn{A21} do not contribute to the classical phase shift.
In order to have a classical contribution in the ultra relativistic (or massless) limit we  need  a term in the amplitude proportional to $s^3$ (up to logarithms)\footnote{We should mention that in the massive case other structures emerge both at the one and at the two-loop level. At one-loop the $\Ord(s^2)$ contribution actually contains a classical piece of the form  (with $D=4$ for concreteness)
$
\operatorname{Re} A_1^{(3)}(s,t) \sim  s^2 \frac{m}{q}  \Rightarrow \delta_1 \sim \frac{G_N s}{\hbar} \frac{G_N m}{b}$. At two loops  we can get classical contributions from an amplitude going like $s^2 m^2 \log q^2$ while an amplitude behaving like $s^3 (m/q) \log q^2$ should also be present in order to accommodate the $\delta_0 \delta_1$ interference. Fortunately, all these terms arrange among themselves and do not interfere with the rest of our argument.}
contributing, in general, to both the real and the imaginary parts of $A_2$.

Let us parametrize the latter in the form
\begin{equation}
\operatorname{Im} A_2^{(2)}(s,t) = G_2(t, m_i^2) s^3  \log^p(s) ~~{\rm with~some}~ p > 0 \, .
\label{ImA22}
\end{equation}
We may then use \eqn{nodd} of Appendix A to get \footnote{In order to keep track of the first sub-leading correction in $m^2/s$, as we did until now, we should include such corrections also in \eqref{ImA22} and \eqref{ReA22}. We leave this point to our forthcoming paper\cite{toap}.}:
\begin{equation}
\operatorname{Re} A_2^{(2)}(s,t) =  \frac{\pi p}{2 \log s} \operatorname{Im} A_2^{(2)}(s,t) \left(1 + \Ord\left(\frac{1}{\log^2 s}\right) \right) \, .
\label{ReA22}
\end{equation}
A highly non-trivial test of \eqn{A21} and \eqn{ReA22} is provided by the explicit results of \cite{Henn:2019rgj}. These are reported, for instance, in Eq.~(B.1) of \cite{DiVecchia:2019kta} where both the structure of \eqn{A21} and the one of \eqn{ImA22} are present. The process discussed there is not necessarily $s-u$ symmetric but becomes so (at the order we consider) by choosing the external states appropriately\footnote{This can be done by choosing an axion and a dilaton as incoming state, which implies that $\hat{A}^{(0)}=(1+\frac{t}{s})+\ldots$ in the notations of~\cite{DiVecchia:2019kta}.}. After doing so  all the real sub-leading terms (of $\Ord(s q^2)$) are matched either by the leading ones on the first line of (B.1) through \eqn{A21}, or by the logarithmically-enhanced imaginary terms through \eqn{ReA22}, or both. Nothing constrains instead the non logarithmically-enhanced imaginary terms at this level since their real counterparts would be sub-sub-leading in $s$.

We finally use exponentiation in impact parameter space to argue that:
 \begin{align}
& \operatorname{Re} \tilde{A}_2(s,b) =\operatorname{Re} \tilde{A}_2^{(1)}(s,b) +\operatorname{Re} \tilde{A}_2^{(2)}(s,b)    =  - \frac43 \delta_0^3 - 4\delta_0 \operatorname{Im} \Delta_1  + 2 \operatorname{Re} \delta_2 \nonumber \\
& \Rightarrow \operatorname{Re} \tilde{A}_2^{(2)}(s,b) = 2 \operatorname{Re} \delta_2 - 2 s^3\widetilde{(t F_2)} - 4 \delta_0 \operatorname{Im} \Delta_1 ~;~ 2 \operatorname{Im} \Delta_1 = - \pi s^2~ \tilde{G}_1 \, ,
\label{RetA22}
\end{align}
 where we used \eqn{A21} including the contribution $2 s^3 t F_2$ to $2 \operatorname{Re}( \delta_2)$. Analogously,
\begin{equation}
\operatorname{Im} \tilde{A}_2(s,b) = 2 \operatorname{Im} \delta_2 + 4 \delta_0 \operatorname{Re}\Delta_1 ~;~ 2 \operatorname{Re}\Delta_1 =  2 s^2 \log s ~\tilde{G}_1 + \frac{3}{\pi} s^2 \log s ~ \widetilde{(t F_1)}\, ,
\label{ImtA22}
\end{equation}
 where the  term $4  \delta_0 \operatorname{Re}(\Delta_1)$ represents the full elastic contribution to the $s$-channel discontinuity of the amplitude while $2  \operatorname{Im} (\delta_2)$ represents the inelastic (3 particle) contribution.
 
 We are interested in extracting $\operatorname{Re}( \delta_2)$ from Eqs.~\eqref{RetA22} and \eqref{ImtA22}. In particular we would like to connect $\operatorname{Re}( \delta_2)$ to $\operatorname{Im}( \delta_2)$, an easier quantity to evaluate.  Since $F_1(t,M_i^2)$ and $F_2(t, m_i^2)$ are both known in terms of $\delta_0$ (see below) the only obstacle for obtaining the sought for relation is to eliminate the non-universal $G_1(t,m_i^2)$ contribution appearing in $\operatorname{Re}(\Delta_1)$ and $\operatorname{Im} (\Delta_1)$. Using \eqn{ReA22} a straightforward calculation gives:
 \begin{align}
& 2 \operatorname{Re} (\delta_2) =  \frac{\pi p}{2 \log s} (2 \operatorname{Im} \delta_2) + \pi (p-1) 2 \delta_0 s^2  \tilde{G}_1  +\frac{3p}{2} s^2 (2 \delta_0) \widetilde{(t F_1)} + 2 s^3 \widetilde{(tF_2)} + \Ord\left(\frac{1}{\log s}\right) \nonumber \\
&=  \frac{\pi p}{2 \log s} (2 \operatorname{Im} \delta_2)   - \frac{4-3p}{s}   \delta_0  (2 \nabla  \delta_0)^2 -  (p-1) (2 \delta_0) (2 \operatorname{Im} \Delta_1) + \Ord\left(\frac{1}{\log s}\right) \, ,
\label{Redelta2} 
 \end{align}
where in the last equation we have used 
 \eqn{delta02} and \eqn{A21} to express $ \widetilde{(t F_1)}$ and $\widetilde{(t F_2)}$ in terms of $\delta_0$, and \eqn{A12} to express $ \tilde{G}_1$ in terms of $\operatorname{Im} \Delta_1$.
 
We thus notice that only for $p=1$ is $\operatorname{Re}(\delta_2)$ given entirely in terms of $\operatorname{Im} \delta_2$ and of $\delta_0$. For $p \ne 1$, $\operatorname{Re}(\delta_2)$ will also depend on $\tilde{G}_1$ i.e. on $\operatorname{Im}( \Delta_1)$ which is non universal. We shall see in Section~\ref{highenergy} that $\operatorname{Im}(\delta_2)$ is indeed universal at high energy, and that  \eqn{ImA22} holds with $p=1$. As a result, also $\operatorname{Re}(\delta_2)$ is universal in the high-energy limit and given by:
 \begin{equation}
 \operatorname{Re}(2\delta_2) =  \frac{\pi}{2 \log s}  \operatorname{Im}(2\delta_2)  -\frac{ \delta_0 }{s} (\nabla 2 \delta_0)^2 + \Ord\left(\frac{1}{\log s}\right) \, .
\label{Redelta2p1}
\end{equation}
Note that both $\operatorname{Im}(\delta_2)$ and $\delta_0  (\nabla  \delta_0)^2$ are IR divergent, but these divergences cancel so that physical observables derived from $\operatorname{Re}(\delta_2)$, such as the deflection angle, are finite.

\section{High energy limit of the 3-particle cut}
\label{highenergy}

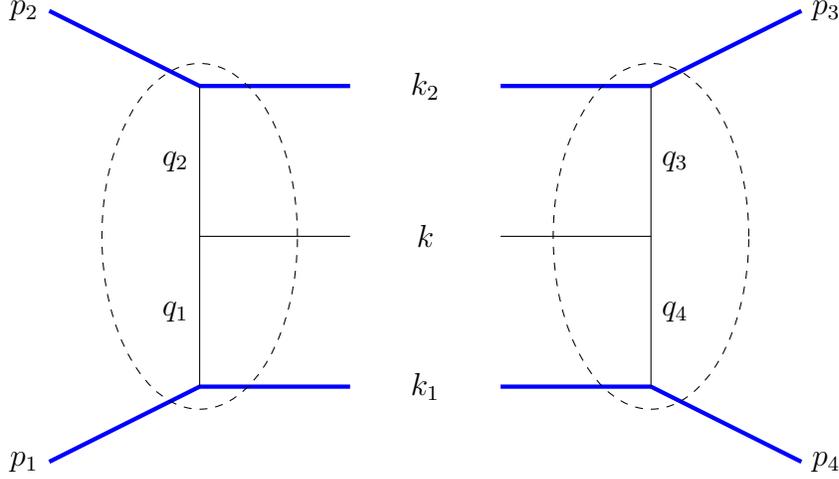
\begin{figure}
\begin{center}
\begin{tikzpicture}
\path [draw, ultra thick, blue] (-5,6)--(-3,5)--(-1,5);
\path [draw, ultra thick, blue] (-5,0)--(-3,1)--(-1,1);
\path [draw] (-3,3)--(-1,3);
\path [draw] (-3,1)--(-3,5);
\path [draw, ultra thick, blue] (5,6)--(3,5)--(1,5);
\path [draw, ultra thick, blue] (5,0)--(3,1)--(1,1);
\path [draw] (3,3)--(1,3);
\path [draw] (3,1)--(3,5);
\draw[dashed] (-3,3) ellipse (1.3 and 2.3);
\draw[dashed] (3,3) ellipse (1.3 and 2.3);
\node at (0,5){$k_2$};
\node at (0,3){$k$};
\node at (0,1){$k_1$};
\node at (-5,6)[left]{$p_2$};
\node at (5,6)[right]{$p_3$};
\node at (-5,0)[left]{$p_1$};
\node at (5,0)[right]{$p_4$};
\node at (-3,4)[left]{$q_2$};
\node at (3,4)[right]{$q_3$};
\node at (-3,2)[left]{$q_1$};
\node at (3,2)[right]{$q_4$};
\end{tikzpicture}
\end{center}
\caption{\label{fig:3pdisc} The lines in bold represent energetic massive states, while the others represent massless states. The process depicted inside the dashed bubbles should not be interpreted as a specific Feynman diagram contribution, but just as a visual aid to recall the definition of the kinematic variables $q_i$. We are interested in the {\em full} $2\to 3$ tree level process, see~\eqref{ampli} as an explicit example.}
\end{figure}

In this section we focus on the 3-particle cut contribution to $A_2$ as depicted in Fig.~\ref{fig:3pdisc}. 
Since we are interested in the ultra-relativistic limit, the amplitude is dominated by the graviton exchange whose contribution is universal. Thus for the sake of simplicity we focus on a gravity theory that can be obtained by taking the double copy of gauge theory amplitudes or equivalently the field theory limit of string amplitudes. The presence of extra fields, and the dilaton in particular, becomes irrelevant in the limit $s\gg m_i^2$. In order to obtain an explicit result we consider a 5-point amplitude in bosonic string theory where the external states $p_1$ and $k_1$ have a Kaluza-Klein momentum along one compact direction and $p_2$ and $k_2$ along another so they describe massive scalars in the uncompact dimensions\footnote{We used the KLT procedure to obtain the closed string amplitude from the open string one, see for instance Eq.~(5.3) of~\cite{BjerrumBohr:2010zs}: in our case, in order to encode the dependence on the KK masses, one needs to use $2 k_i k_j$, instead of $s_{ij}$, in the second line of that equation where all the scalar products are restricted to the uncompact directions.}. By taking the field theory limit we obtain the following result for the $2\to 3$ process in the left part of Fig.~\ref{fig:3pdisc},
\begin{align}
  \label{ampli}
M^{\mu \nu}  = &\;  2(8\pi G_N)^{\frac{3}{2}} \Bigg\{  (k_1p_2)(k_2 p_1) \left( -\frac{k_{1\mu}}{k_1 k}  +  \frac{k_{2\mu}}{k_2 k} \right)\left( -\frac{p_{2\nu}}{p_2 k}  +  \frac{p_{1\nu}}{p_1 k}  \right) + 4 q_1^2 q_2^2 \\
\times &   \Bigg[ \frac{q_1^\mu (p_1p_2) + p_2^\mu (p_1k) - p_1^\mu (p_2k)}{q_1^2 q_2^2} 
 +  \frac{k_2^\mu}{2k_2 k} \left( \frac{p_1p_2}{q_1^2}  + \frac{1}{2} \right) - \frac{k_1^\mu}{2k_1k} \left( \frac{p_1p_2}{q_2^2} +\frac{1}{2} \right) \Bigg] \nonumber \\
 \times & \Bigg[ \frac{q_1^\nu (k_1k_2) + k_2^\nu (k_1k) - k_1^\nu (k_2k) }{q_1^2 q_2^2} 
 - \frac{p_1^\nu}{2p_1 k} \left( \frac{k_1k_2}{q_2^2} + \frac{1}{2} \right) + \frac{p_2^\nu}{2p_2 k} \left( \frac{k_1k_2}{q_1^2} + \frac{1}{2} \right) \Bigg]  \Bigg\},
\nonumber
\end{align}
where we introduced the variables $q_1 =-p_1-k_1$, $q_2 = - p_2-k_2$, which satisfy $k=q_1+q_2$. The overall normalization was fixed by considering the leading high energy term for the Weinberg soft limit $k\to 0$, where~\eqref{ampli} has to reduce to the leading $2\to 2$ amplitude times a universal factor.

One can then  calculate the contribution to the imaginary part of $A_2$ from the 3-particle cut by the usual phase space integral
\begin{align}
  2 \operatorname{Im} A_2^{(3p)}  & = \frac{1}{\left((2\pi)^{D-1}\right)^3} \int \frac{d^{D-1} k_1}{2E_{k_{1}}}\int \frac{d^{D-1} k_2}{2E_{k_{2}}} \int \frac{d^{D-1} k}{2E_{k}} 
  {\it M}^{\mu \nu}(p_1, p_2 ; k_1, k_2 ,k)  \nonumber \\
 \times & 
{\it M}_{\mu \nu} (-k_1,-k, -k_2; p_3, p_4) (2\pi)^{D} \delta^{(D)} (p_1+p_2 + k_1 +k_2 +k)\;.
\label{UR1}
\end{align}
Here we are interested in the ultra-relativistic limit, so it is possible to approximate~\eqref{UR1} in the double Regge limit\footnote{This limit can be performed on the full $2\to 3$ amplitude~\eqref{ampli} by scaling $k_1^\mu$ and $k_2^\mu$ as $s_{1,2}$ and one obtains directly Eq.~(4.10) of~\cite{Ademollo:1990sd}; notice that the latter result is traceless showing explicitly that the dilaton decouples in the double Regge limit.}
\begin{equation}
  \label{eq:DRegge}
  s\gg s_1, s_2 \to \infty\;, \quad \mbox{with }~ \frac{s_1 s_2}{s}\;,~ q_i^2\;,~ m_i^2~\mbox{ fixed}\;,
\end{equation}
where $s=-(p_1+p_2)^2$ and $s_i= - (k+k_i)^2$. In this regime it is convenient to write the kinematic variables in terms of the $(D-2)$ space-like vectors orthogonal to the direction where the energetic states are boosted (which we take to be $x^{D-1}$). By working in the Breit frame and taking light-cone variables for the time and longitudinal direction $(p_0 +p_{D-1},\vec{p}, p_0-p_{D-1})$, we have
\begin{align}
p_1 & = ( {\overline{m}}_1 {\rm e}^{y_1}, -\frac{\vec{q}}{2}, {\overline{m}}_1 {\rm e}^{-y_1})~;~p_2= ( {\overline{m}}_2 {\rm e}^{y_2}, \frac{\vec{q}}{2}, {\overline{m}}_2 {\rm e}^{-y_2}) \nonumber \\
p_4 & = (- {\overline{m}}_1 {\rm e}^{y_1}, -\frac{\vec{q}}{2}, -{\overline{m}}_1 {\rm e}^{-y_1})~~;~~p_3 = (-{\overline{m}}_2 {\rm e}^{y_2}, \frac{\vec{q}}{2}, {-\overline{m}}_2 {\rm e}^{-y_2})\;,
\label{PHI10}
\end{align}
where $y_i$ are the rapidities of the external particle and
${\overline{m}}_{1,2}^2 =m_{1,2}^2 + \frac{\vec{q}^{\;2}}{4}$.
The intermediate states with momentum $k_1, k_2, k$ (all incoming) are given by
\begin{align}
k_1  & = ( -{\overline{m}}_1 ' {\rm e}^{y_1'},  \frac{\vec{q}}{2}-{\vec{q}}_1, -{\overline{m}}_1 ' {\rm e}^{-y_1'})~~;~~
k_2 = (-{\overline{m}}_2 ' {\rm e}^{y_2'}, -\frac{\vec{q}}{2}-{\vec{q}}_2, {-\overline{m}}_2 ' {\rm e}^{-y_2'})\;,
\nonumber \\
k & = (-|k| {\rm e}^{y}, \vec{k} , -|k| {\rm e}^{-y})\;,
\label{PHI12}
\end{align}
 where
$({\overline{m}}_{1} ')^2 = m_1^2 + ( \frac{\vec{q}}{2} -{\vec{q}}_1)^2$ and $({\overline{m}}_{2} ')^2 = m_2^2 + ( \frac{\vec{q}}{2} +{\vec{q}}_2)^2$. 
In the double Regge limit, one can use approximations such as $q_i^2 \sim \vec{q}_i^{\;2}$ and~\eqref{ampli} reduces to the result of~\cite{Amati:1990xe} and Eq.~(4.10) of~\cite{Ademollo:1990sd} even if in those papers the external states are massless. This is not surprising since we are keeping the masses $m_i^2$ fixed as the Mandelstam variables $s$, $s_i$ become large, however there is a point that deserves a further comment. By always keeping the leading order when rewriting the $D$-dimensional kinematics in terms of the transverse, we obtain that the ultra-relativistic limit of Eq.~\eqref{UR1} agrees with the result of~\cite{Amati:1990xe}
\begin{align} \label{fqi}
\operatorname{Im} A_2^{(3p)}  \simeq &~ \frac{(16 \pi G_N)^3s^3}{2\pi} \int dy  \int \frac{d^{D-2} \vec{q}_1}{(2\pi)^{D-2}}  \int \frac{d^{D-2} \vec{q}_2}{(2\pi)^{D-2}}  \frac{1}{(\vec{k}^2)^2}  \\
 \times & \left[  \frac{\left[  (\vec{q}_1 \vec{q}_4)(\vec{q}_2 \vec{q}_3) + (\vec{q}_1 \vec{q}_2)(\vec{q}_3 \vec{q}_4) -
(\vec{q}_1 \vec{q}_3)(\vec{q}_2 \vec{q}_4) \right]^2}{\vec{q}_1^{\;2} \vec{q}_2^{\;2} \vec{q}_3^{\;2} \vec{q}_4^{\;2}}   +1  -  \frac{( \vec{q}_1 \vec{q}_2)^2}{\vec{q}_1^{\;2} \vec{q}_2^{\;2}}-  \frac{(\vec{q}_3 \vec{q}_4)^2}{\vec{q}_3^{\;2} \vec{q}_4^{\;2}} \right], \nonumber
\end{align}
where we used the delta function in~\eqref{UR1} in order to perform the integrals over the rapidities\footnote{This is possible only if $m_{t,k}{\rm e}^{\pm y} \leq \sqrt{s}$, which provides the limits of integration for the rapidity variable $y$.} $y'_{1,2}$ and over the spatial components $\vec{k}$. This result has a milder than naively expected IR behaviour since in~\eqref{UR1} there are no terms diverging as ${q}_1^{-2} {q}_2^{-2}$ (or as  ${q}_3^{-2} {q}_4^{-2}$) the the relevant ${q}_i$ are small. The cancellation of such contribution ensures that there are no $1/\epsilon^2$ contributions in the massless case and, crucially for us, no $\log^2(s)$-enhanced terms, {\em i.e.} no contribution with $p=2$ in~\eqref{ImA22}. This can be seen as follows: terms where the integration over the $\vec{q}_i$ are factorised can produce $1/\epsilon^2$ contribution if each integration is independently IR divergent\footnote{The UV divergences in~\eqref{ampli} cancel, see~\cite{Amati:1990xe}.}, however in this case, one should keep subleading corrections to the approximation $q_i^2 \sim \vec{q}_i^{\;2}$ by including terms that break the factorisation such as $q_1^2 = \vec{q}_1^{\;2} + A (\vec{q}_1+\vec{q}_2)^2 +\ldots$ and $q_2^2 = \vec{q}_2^{\;2} + B (\vec{q}_1+\vec{q}_2)^2 + \ldots$; then $A B\sim (m_1 m_2/s)^2$ acts as regulator in the deep IR region producing a contribution proportional to $\epsilon^{-1} \log(m_1 m_2/s)$ instead of $1/\epsilon^2$. Since terms diverging as ${q}_1^{-2} {q}_2^{-2}$ are absent in~\eqref{ampli} then this mechanism does not apply and the only possibility to generate a factor of $\log(s)$ is from the integration over $y$.

Thus starting from~\eqref{fqi} it is possible to follow the derivation of~\cite{Amati:1990xe} and obtain
\begin{equation}
  \operatorname{Im}  {\widetilde{A_2^{(3p)}}} (s, b) \simeq \frac{1}{2 s} \, \frac{ (8G_N s)^3  \log s \Gamma^3 (1-\epsilon)}{16 (\pi b^2)^{1-3\epsilon}}
\left[ - \frac{1}{4\epsilon}+ \frac{1}{2} + {\cal{O}} (\epsilon) \right]\;.
\label{expaMtilde}
\end{equation}
This result is nothing else than $2\operatorname{Im}(\delta_2)$ and we can use it in the general result~\eqref{Redelta2p1}. First one can check that, as in \cite{Amati:1990xe} for the massless case, IR divergences cancel in \eqn{Redelta2p1} and notice that for this cancellation to happen it is crucial that~\eqref{expaMtilde} implies that $p=1$ in~\eqref{ImA22}. Then the finite term provides
finite, smooth, and universal result for the high-energy limit of $\operatorname{Re}(\delta_2)$ in $D=4$:
 \beq\label{eq:ur}
 \operatorname{Re}(2\delta_2)  \simeq \frac{4 G_N^3 s^2}{\hbar b^2}
  \eeq
in agreement with \cite{Amati:1990xe,DiVecchia:2019kta,Bern:2020gjj}.

\section{Direct calculation of $\operatorname{Re} \delta_2$}
\label{direct}

In order to corroborate with an explicit example the general results obtained in the previous sections, in this section we provide an explicit expression for the four-point two-loop amplitude in ${\cal{N}}=8$ supergravity for the scattering of two scalar particles with masses $m_1$ and $m_2$. This case has been already studied at one loop in~\cite{Caron-Huot:2018ape} and at two loops in~\cite{Parra-Martinez:2020dzs}. We follow closely the procedure of~\cite{Parra-Martinez:2020dzs} where the two-loop amplitude in ${\cal{N}}=8$ massive supergravity is written in terms of a set of basic scalar integrals $I_\mathrm{T}$, where the subscript $\mathrm{T}\in\{\mathrm{III},\overline{\mathrm{III}},\mathrm{IX},\overline{\mathrm{IX}}\}$ indicates the diagram's topology,
\begin{align}
A_2 (s, q^2) = & ~ \frac{(8\pi G_N)^3}{2} \Bigg( (s-m_1^2-m_2^2)^4 + (u-m_1^2-m_2^2)^4 - t^4 \Bigg)
\nonumber \\
\times & \Bigg[(s-m_1^2-m_2^2)^2 \left( I_{\rm III} + I_{\rm IX} + I_{\rm XI}\right) \nonumber \\
 +&  (u-m_1^2-m_2^2)^2 \left( I_{\overline{\rm III}} +I_{\overline{\rm IX}} +I_{\overline{\rm XI}} \right) +t^2 \left( I_{\rm H} + I_{\overline{\rm H}} + \cdots \right)\Bigg]\,.
\label{CR1}
\end{align}
Here we follow the notation of Eq.~(3.16) of~\cite{Parra-Martinez:2020dzs} and, for the sake of simplicity, we have taken the angle appearing in that reference specifying the Kaluza-Klein reduction to be $\phi = \frac{\pi}{2}$; in addition we have chosen the incoming particles to be an axion and a dilaton to give an amplitude that is symmetric under the exchange of $s$ with $u$. Finally we neglected all the integral structures that are subleading in the high energy regime we are interested in or, equivalently, in limit of small momentum transfer.

The integrals $I_{\rm H}$ and $I_{\overline{\rm H}}$ were computed in an $\epsilon$ expansion for arbitrary kinematics in~\cite{Bianchi:2016yiq}, while the remaining integrals in~\eqref{CR1} were studied in~\cite{Parra-Martinez:2020dzs}: they adapted the differential equation approach, adopted in particular in \cite{Henn:2013woa} for the double box integral\footnote{The double box and non-planar double box were also calculated in \cite{Smirnov:2001cm,Heinrich:2004iq,Henn:2013woa} via a Mellin--Barnes representation; whenever possible we checked that the results in our 
Appendix~\ref{Appb} are consistent with the papers mentioned above.}, by implementing the soft limit $|t| \ll s, m_1^2, m_2^2$ from the beginning and then further simplified the problem by calculating the boundary conditions of the relevant differential equation in the ``potential'' approximation. Here instead we do not take this extra approximation and we provide the result valid in the full soft region even if we focus on the ultra-relativistic $(s\gg m_i^2)$ case. We collect in Appendix~\ref{Appb} the results for all scalar integrals needed and leave a detailed discussion of the generic kinematics to a followup paper~\cite{toap}.

When we add all the contributions we see that all $\log (s)$ in the real part of the amplitude cancel and one is left only with a $\log(s)$ in the imaginary part. In conclusion the complete ultra-relativistic amplitude is
 \begin{equation}
 \label{CR27}
 \begin{split}
A_2 (s, q^2) & \simeq \frac{(8\pi G_N)^3 s^3}{(4\pi)^4}  \Bigg\{ - \frac{2\pi^2 s}{\epsilon^2 q^2}  \left( \frac{4\pi {\rm e}^{-\gamma_E}}{q^2}\right)^{2\epsilon}   - \frac{4\pi (i -\pi)}{\epsilon^2}\left( \frac{4\pi {\rm e}^{-\gamma_E}}{q^2}\right)^{2\epsilon}   \\
&+  \frac{1}{\epsilon}\left( \frac{4\pi {\rm e}^{-\gamma_E}}{q^2}\right)^{2\epsilon}  \left[ 4\pi^2 + 8\pi i \log \frac{s}{m_1m_2} -8\pi i - i \frac{\pi^3}{3}  \right] \Bigg\} + {\cal{O}}(\epsilon^0)\,.~
\end{split}
\end{equation} 
Before proceeding further, let us notice that this amplitude perfectly satisfies some of the general properties discussed in Sect.~\ref{analyticity}. In fact,  the ratio between the leading and the real part of the subleading term at the order $\frac{1}{\epsilon^2}$ is equal to $\frac{s}{2t}$ in agreement with the first line of Eq.~\eqref{A21}, while, at the order $\frac{1}{\epsilon}$,  the first two terms of the square bracket satisfy Eq.~\eqref{ReA22} for $p=1$. 
This is the consequence of a non-trivial cancellation since the individual integrals contain higher $\log^p(s)$ contributions, see for instance the contribution coming from $I_{\rm H}$ in~\eqref{HHbar}. There are also further non-trivial cancellations that ensure that the leading term at large distance, proportional to $(q^2)^{-1+2\epsilon}$, takes a particularly simple form in line with the expectation of the eikonal exponentiation~\eqref{eq:25}. This is more easily seen in impact parameter space where we get
\begin{equation}
\label{CR28a}
\begin{split}
{\tilde{A}}_2 (s, b) \simeq &~ \Bigg\{ \frac{ G_N^3 s^3 (\pi b^2 )^{3\epsilon}\Gamma^3 (1-\epsilon)}{6   \epsilon^3} - \frac{8G_N^3 (i -\pi)s^2 (\pi b^2 )^{3\epsilon}\Gamma^3 (1-\epsilon) }{\epsilon  \pi b^2   }  \\
+ & \frac{2 G_N^3s^2 \Gamma^3 (1-\epsilon)}{(\pi b^2)^{1-3\epsilon} } \left[ 4\pi + 8 i \log \frac{s}{m_1m_2} -8 i - i \frac{\pi^2}{3}  \right] + \Ord(\epsilon) \Bigg\}\;.
\end{split}
\end{equation}

In order to compute the new contribution to the eikonal we must first subtract the contribution of the lower eikonal $\delta_0$ and $\Delta_1$ that are equal to\footnote{The result for $\operatorname{Im}(2 \Delta_1)$ is apparently different from the one obtained in~\cite{DiVecchia:2019kta} because of the different conventions mentioned in footnote around Eq.~\eqref{eq:25}}
\begin{eqnarray}
2\delta_0 \simeq  \frac{G_N s \Gamma (1-\epsilon) (\pi b^2)^{\epsilon} }{-\epsilon}~~;~~2 \operatorname{Im} \Delta_1 \simeq 
 \frac{8 G_N^2 s (\pi b^2)^{2\epsilon} \Gamma^2 (1-\epsilon)}{b^2} \left(1+ \frac{\epsilon}{2} \right)\;.
\end{eqnarray}
Using them we can write the real part of the amplitude in Eq. \eqref{CR28a} as follows:
\begin{equation}
\operatorname{Re} {\tilde{A}}_2 (s, b) \simeq  - \frac{i}{6}(2i\delta_0)^3 - \operatorname{Im} (2\Delta_1) 2 \delta_0 +\frac{4G_N^3 s^2 (\pi b^2)^{3\epsilon} \Gamma^3 (1-\epsilon)}{ b^2} + \Ord(\epsilon)\,.~~~~~~
\end{equation}
The last term in this equation is the ultra-relativistic limit of $\operatorname{Re}(\delta_2)$ and is immediate to check that in $D=4$ one recovers the universal result~\eqref{eq:ur} obtained in the  previous section by generalising the approach of~\cite{Amati:1990xe}.

For completeness, let us also present the real part of the eikonal to third post-Minkowskian order for generic $s=m_1^2+m_2^2+2m_1m_2\sigma$ (with $\sigma \geq1$)~\cite{toap},
\begin{equation}\label{}
  \operatorname{Re}(\delta_2)= \frac{2 G_N^3  (2m_1 m_2 \sigma)^2 }{b^2}\left[\frac{\sigma^4}{
\left(\sigma
^2-1\right)^2}-\cosh ^{-1}(\sigma ) \left(\frac{\sigma^2}{\sigma ^2-1}-\frac{\sigma ^3
\left(\sigma ^2-2\right)
}{\left(\sigma ^2-1\right)^{5/2}}\right) \right]\,.
\end{equation}
Furthermore, the corresponding 3PM contribution to the scattering angle as a function of the angular momentum $J$ reads 
\begin{align}\label{eq:chi3}
\chi_{3\mathrm{PM}}=&-\frac{16 m_1^3 m_2^3 \sigma ^6 G_N^3}{3 J^3
	\left(\sigma ^2-1\right)^{3/2}}+\frac{32 m_1^4 m_2^4
	\sigma ^6 G_N^3}{J^3 \left(\sigma ^2-1\right)s}
\\\nonumber
&- \frac{4}{J^3 s}  \left(16 m_1^4 m_2^4 \sigma ^4 G_N^3-\frac{16 m_1^4 m_2^4 \sigma^5 \left(\sigma ^2-2\right) G_N^3}{\left(\sigma ^2-1\right)^{3/2}}\right) {\rm arcsinh}\sqrt{\frac{\sigma-1}{2}}\;.
\end{align}
Let us conclude with a few comments on the 3PM result~\eqref{eq:chi3}\footnote{The apparent contradiction between~\eqref{eq:chi3} and the results of~\cite{Parra-Martinez:2020dzs} can be attributed to the fact that the latter paper, as well as~\cite{Bern:2019nnu, Bern:2019crd},  compute an unphysical ``conservative scattering angle" (to be used for determining the conservative part of the EOB potential) while we are dealing with the physical deflection angle including radiation-reaction effects. We thank Zvi Bern for this important observation.}. The first term in the first line is entirely determined by $\chi_{1\mathrm{PM}}$, while the first term in the second line is due to integrals $I_{\rm H}$ and $I_{\overline{\rm H}}$, see~\eqref{HHbar}. Together they reproduce exactly Eq.~(6.41) of~\cite{Parra-Martinez:2020dzs}. The second term in the second line is a contribution coming from the full soft-region analysis of the crossed double-box integrals. In the ultra-relativistic limit $\sigma \gg 1$ the leading term ${\cal O}(\sigma^4$) in each term in the round parenthesis in the second line cancels and only the first line survives reproducing the universal and finite ultra-relativistic result which was the main focus of this paper. Thanks to analyticity/crossing argument in Section~\ref{analyticity}, this cancellation is a consequence of the cancellation in the imaginary part mentioned below Eq.~\eqref{CR27}. Notice that the functional form of~\eqref{eq:chi3} matches exactly Eq.~(3.65) of~\cite{Damour:2019lcq} (where of course the Schwarzschild contribution is substituted by the probe-limit relevant to this supersymmetric case) and so it is possible to define a function $\overline{C}(\sigma)$ also for the ${\cal N}=8$ case.

It is interesting to look at the opposite limit $\sigma \to 1$ which is relevant to the PN (Post-Newtonian) limit: the terms already present in Eq.~(6.41) of~\cite{Parra-Martinez:2020dzs} have a standard $n$PN expansion where $n$ is integer. The new terms yield contributions only at half-integer PN orders, starting at $1.5$PN, and so they do not modify the integer PN data. However, these half-integer PN terms, usually associated with dissipative phenomena, are somehow unexpected at order $G_N^3$, so it would certainly be interesting to repeat the same analysis for pure GR, instead of ${\cal N}=8$ supergravity case. If a similar pattern appears also in GR, as it seems reasonable since the same integrals $I_{\rm T}$ appear in all cases, that would of course be the best setup where to investigate these issues   in more detail.

\vspace{5mm}

\noindent {\large \textbf{Acknowledgements} }

We thank Zvi Bern, Arnau Koemans Collado, Thibault Damour, Stephen Naculich, Julio Parra Martinez, Augusto Sagnotti and Congkao Wen for useful discussions. We thank Claude Duhr and Vladimir Smirnov for helping us check with independent methods some of our results and Gudrun Heinrich for some numerical checks. The research of RR is partially supported by the UK Science and Technology Facilities Council (STFC) Consolidated Grant ST/P000754/1 ``String theory, gauge theory and duality''. The research of CH (PDV) is fully (partially) supported by the Knut and Alice Wallenberg Foundation under grant KAW 2018.0116. PDV, RR and GV would like to thank the Galileo Galilei Institute for hospitality during the workshop ``String theory from the worldsheet perspective'' where they started discussing this topic.

\appendix

\section{Asymptotic $\rho \equiv \frac{\operatorname{Re} A}{\operatorname{Im} A}$ from analyticity and crossing}
\label{Appa}

Consider a crossing-symmetric (under $s \leftrightarrow u$), real-analytic amplitude $A(s^\ast, t) = A^\ast(s, t)$ as a function of the complex variable $s$ at $t \le 0$ and  assume that, as $s \to \infty$,
\beq
\operatorname{Im} A(s,t) \sim s^n \log^p s ~~,~~  |A| s^{-n-1} \to 0 ~~;~~   (n, p \ge  0)
\label{as}
\eeq
Since real-analyticity and crossing symmetry imply $\operatorname{Im} A(-s + i 0) = - \operatorname{Im} A(s + i 0)$, we can  write an ($n+1$)-times subtracted dispersion relation in the form:
\beq
\operatorname{Re} A(s,t) = Q_{2m}(s,t) + \frac{2 }{\pi} s^{2m+2}~ {\cal P} \int_{s_0}^{\infty} ds' \frac{\operatorname{Im}  A(s',t)}{s'^{2m+1} (s'^2- s^2)} 
\label{SDR}
\eeq
where $s_0=(m_1+m_2)^2$ is the $s$-channel threshold, $ Q_{2m}(s,t)$ is a polynomial of degree $2m$, ${\cal P}$ denotes the principal part, and the integer $m$ is defined by $n=2 m$ for $n$ even or by $n=2m +1$ for $n$ odd\footnote{We have neglected corrections from the finite-$s,u$ regions since they give sub-leading contributions similar to those coming from $ Q_{2m}(s,t)$.}. Because of \eqn{as} the integral in \eqn{SDR} converges.
From \eqn{SDR} we can compute $\rho$ distinguishing the two cases.

For $n$ even we get:
\begin{eqnarray}
&&  \frac{\operatorname{Re} A(s,t)}{s^{n}} = \frac{2 }{\pi} s^{2} {\cal P}  \int_{s_0}^{\infty} ds' \frac{\operatorname{Im}  A(s',t) s'^{-n}}{s'(s'^2- s^2)} + \frac{Q_{n}(s,t)}{s^n} \nonumber \\
&& \Rightarrow \rho = \frac{2 }{\pi} s^{2} (\log s)^{-p} {\cal P}  \int_{s_0}^{\infty} ds' \frac{\log^p s'}{s'(s'^2- s^2)} + \frac{Q_{n}(s,t)}{s^n \log^p s} \sim - \frac{2 \log s}{(1+p)\pi}  \; ,
\label{neven}
\end{eqnarray}
while for $n$ odd:
\begin{eqnarray}
&&  \frac{\operatorname{Re} A(s,t)}{s^{n}} = \frac{2 }{\pi} s {\cal P}  \int_{s_0}^{\infty} ds' \frac{\operatorname{Im}  A(s',t) s'^{-n}}{(s'^2- s^2)} + \frac{Q_{n-1}(s,t)}{s^n} \nonumber \\
&& \Rightarrow \rho = \frac{2 }{\pi} s (\log s)^{-p} {\cal P}  \int_{s_0}^{\infty} ds' \frac{\log^p s'}{(s'^2- s^2)} + \frac{Q_{n-1}(s,t)}{s^n \log^p s} \sim \frac{\pi p}{2  \log s } \; ,
\label{nodd}
\end{eqnarray}
where we have used the high-energy limit of the principal-part integrals appearing in \eqn{neven} and \eqn{nodd}. These are explicitly known in terms of special functions.

An exceptional case (which we encounter for the amplitude $A_1^{(1)}$ of \eqn{A11}) is the one of \eqn{nodd} with $p=0$. In that case $\rho$ is suppressed by a full power of s but the leading term comes from picking up the first subleading $n$-even correction in \eqn{as}, which is fixed by crossing symmetry, and from applying to it \eqn{neven}.
The last case needed, the one of $A_2^{(1)}$, is instead trivial since the amplitude is purely real and crossing symmetry forces the combination $s^4 + 2 t s^3 + \Ord(t^2 s^2)$.

\section{Summary of the soft integrals}
\label{Appb}

In this appendix, we collect the expressions for the scalar integrals that enter the $\mathcal N=8$ two-loop amplitude.
We focus on the contributions that are non-analytic in $q^2$, which emerge from the soft region in the small-$q$ expansion.
In order to exhibit the relevant expressions for each topology $\mathrm{T}\in\{\mathrm{III},\overline{\mathrm{III}},\mathrm{IX},\overline{\mathrm{IX}}\}$, let us denote
\begin{equation}
\begin{aligned}
I_{\mathrm{T}}=
\frac{1}{(4\pi)^4}\left( \frac{4\pi e^{-\gamma_E}}{q^2} \right)^{2\epsilon}
\left(
\frac{I_{\mathrm{T}}^{(2,2)}}{\epsilon^2 q^2}
+
\frac{I_{\mathrm{T}}^{(1,2)}}{\epsilon\, q^2}
+
\frac{I_{\mathrm{T}}^{(2,0)}}{\epsilon^2}
+
\frac{I_{\mathrm{T}}^{(1,0)}}{\epsilon} + \cdots
\right)\,.
\end{aligned}
\end{equation}
Here the dots stand for terms that are subleading in $\epsilon$ or in $q$, and we omit terms proportional to $\frac{1}{q}$ which are not needed to extract $\delta_2$ from the amplitude but only to check the $\delta_0 \delta_1$ term arising from the exponentiation.
Let us also introduce the rescaled Mandelstam variable
\begin{equation}\label{key}
\tilde s = \frac{s}{m_1 m_2}\,.
\end{equation}
Then one has, up to subleading orders for $s\gg m_i^2$,
\begin{align}
&I_{\mathrm{III}}^{(2,2)} \simeq
\frac{2 \left(\log \tilde s-i\pi\right){}^2}{s^2}\,,  ~~
I_{\mathrm{III}}^{(2,0)} \simeq 0\,,\\
&I_{\mathrm{III}}^{(1,2)}\simeq
\frac{1}{s^2} \left(\frac{2}{3} \log ^3\tilde s-2 i \pi  \log
	^2\tilde s-\frac{5}{3} \pi ^2 \log \tilde s-2 \zeta_3
	+\frac{i \pi ^3}{3} \right)\,,
\\
&I_{\mathrm{III}}^{(1,0)} \simeq 
\frac{\left(m_1^2+m_2^2\right) \left(\log \tilde s-i \pi \right)}{2
	m_1^2 m_2^2 s^2}
\\ \nonumber
&+\frac{1}{s^3} \left(-2 \log
	^2\tilde s+\left(\tfrac{m_1^2}{m_2^2}+\tfrac{m_2^2}{m_1^2}+4 i \pi
	+4\right) \log \tilde s-\frac{i (2 \pi -i)}{2}
	\left(\tfrac{m_1^2}{m_2^2}+\tfrac{m_2^2}{m_1^2}\right)+2 \pi ^2  -4 \pi i-1 \right),
\end{align}

\begin{align}
&I_{\overline{\mathrm{III}}}^{(2,2)} \simeq
\frac{2 \log ^2\tilde s}{s^2}\,,~~ I_{\overline{\mathrm{III}}}^{(2,0)} \simeq \frac{4 \left(\log \tilde s-1\right) \log \tilde s}{s^3}\,,\\
&I_{\overline{\mathrm{III}}}^{(1,2)}\simeq\frac{2 \log ^3\tilde s+\pi ^2 \log \tilde s-6 \zeta_3}{3
	s^2}\,,\\
&I_{\overline{\mathrm{III}}}^{(1,0)} \simeq \frac{\left(m_1^2+m_2^2\right) \log \tilde s}{2m_1^2m_2^2
	s^2}\\
\nonumber
&+ \frac{1}{6 s^3} \left(8 \log ^3\tilde s+4 \pi ^2 \log \tilde s+\frac{3
		\left(m_1^4+m_2^4\right) \left(2 \log \tilde s-1\right)}{m_1^2
		m_2^2}-2 \left(12 \zeta_3+3+\pi ^2\right)\right)\,,
\end{align}

\begin{align}
&I_{\mathrm{IX}}^{(2,2)} \simeq
-\frac{\left(\log \tilde s-i\pi\right) \log \tilde s}{s^2}\,,~~I_{\mathrm{IX}}^{(2,0)} \simeq
-\frac{\left(\log \tilde s-1\right) \left(\log \tilde s-i \pi
	\right)}{s^3}\,,\\
&I_{\mathrm{IX}}^{(1,2)}\simeq
\frac{1}{s^2} \left(-\frac{1}{3} \log ^3\tilde s+2 i \pi  \log
	^2\tilde s+\frac{11}{6} \pi ^2 \log \tilde s+\zeta_3
	-\frac{i \pi ^3}{3} \right)\,,\\
&I_{\mathrm{IX}}^{(1,0)} \simeq 
\frac{i \left(m_1^2+m_2^2\right) \left(2 \pi +i \log
	\tilde s\right)}{4 m_1^2 m_2^2 s^2}
+\frac{1}{s^3} \left(\frac{1}{6} \log \tilde s \left(2 \left(\log
	\tilde s-12\right) \log \tilde s+\pi  (\pi +24
	i)+12\right)\right) \nonumber \\
& +\frac{1}{s^3}\left(\frac{\left(m_1^4+m_2^4\right) \left(-2 \log
		\tilde s+4 i \pi +1\right)}{4 m_1^2 m_2^2}-\zeta_3-i \pi
	+\frac{1}{2} \right)\,,
\end{align}

\begin{align}
&I_{\overline{\mathrm{IX}}}^{(2,2)}\simeq
-\frac{\left(\log \tilde s - i\pi\right) \log \tilde s}{s^2}\,, ~~ I_{\overline{\mathrm{IX}}}^{(2,0)}\simeq-\frac{\left(\log \tilde s-1\right) \left(\log \tilde s-i \pi
	\right)}{s^3}\,,\\
&I_{\overline{\mathrm{IX}}}^{(1,2)}\simeq\frac{1}{s^2} \left(-\frac{1}{3} \log ^3\left(\tilde s\right)-i \pi  \log
	^2\left(\tilde s\right)-\frac{7}{6} \pi ^2 \log \left(\tilde s\right)+\zeta
	(3)+\frac{i \pi ^3}{6} \right)\,,\\
	\nonumber
&I_{\overline{\mathrm{IX}}}^{(1,0)}\simeq
-\frac{\left(m_1^2+m_2^2\right) \left(\log
	\tilde s+i\pi\right)}{4 m_1^2 m_2^2 s^2}
\\\nonumber&	+\frac{1}{12 s^3} \left(\frac{3 (1-2 i \pi )
		\left(m_1^4+m_2^4\right)}{m_1^2 m_2^2}+36 \zeta_3+ 2 \pi (7 \pi - 6 i + i \pi^2) + 6 \right) \nonumber \\
& +\frac{1}{s^3} \left(-\log ^3\tilde s+(5-i \pi ) \log ^2\tilde s+\frac{1}{2}
	\left(-\frac{m_1^4+m_2^4}{m_1^2 m_2^2}-\pi  (3 \pi +4 i)-8\right) \log
	\tilde s \right)\,.
\end{align}
Finally the contribution of the H diagram and of its crossed counterpart is equal to
\begin{align}\label{HHbar}
I_{\mathrm{H}} &\simeq \frac{1}{\epsilon (4\pi)^4} \left( \frac{4\pi {\rm e}^{-\gamma_E} }{q^2}\right)^{2\epsilon}  \frac{2 \log \tilde s \left(\log ^2 \tilde s-3 i \pi  \log
	\tilde s-2 \pi ^2\right)}{3 q^4 s  }\,,\\
I_{\overline{\mathrm{H}}}&\simeq-\frac{1}{\epsilon (4\pi)^4} \left( \frac{4\pi {\rm e}^{-\gamma_E} }{q^2}\right)^{2\epsilon} \frac{2 \log \tilde s \left(\log ^2 \tilde s+\pi ^2\right)}{3
	q^4 s  }\,.
\end{align}

\providecommand{\href}[2]{#2}\begingroup\raggedright\endgroup


\begin{thebibliography}{10}

\bibitem{tHooft:1987vrq}
G.~'t~Hooft, ``{Graviton Dominance in Ultrahigh-Energy Scattering},''
\href{http://dx.doi.org/10.1016/0370-2693(87)90159-6}{{\em Phys. Lett.} {\bf
  B198} (1987)  61--63}.

\bibitem{Amati:1987wq}
D.~Amati, M.~Ciafaloni, and G.~Veneziano, ``{Superstring Collisions at
  Planckian Energies},''
\href{http://dx.doi.org/10.1016/0370-2693(87)90346-7}{{\em Phys. Lett.} {\bf
  B197} (1987)  81}.

\bibitem{Amati:1987uf}
D.~Amati, M.~Ciafaloni, and G.~Veneziano, ``{Classical and Quantum Gravity
  Effects from Planckian Energy Superstring Collisions},''
\href{http://dx.doi.org/10.1142/S0217751X88000710}{{\em Int. J. Mod. Phys.}
  {\bf A3} (1988)  1615--1661}.

\bibitem{Muzinich:1987in}
I.~J. Muzinich and M.~Soldate, ``{High-Energy Unitarity of Gravitation and
  Strings},''
\href{http://dx.doi.org/10.1103/PhysRevD.37.359}{{\em Phys. Rev.} {\bf D37}
  (1988)  359}.

\bibitem{Sundborg:1988tb}
B.~Sundborg, ``{High-energy asymptotics: the one loop string amplitude and
  resummation},''
\href{http://dx.doi.org/10.1016/0550-3213(88)90014-4}{{\em Nucl. Phys.} {\bf
  B306} (1988)  545--566}.

\bibitem{Amati:1990xe}
D.~Amati, M.~Ciafaloni, and G.~Veneziano, ``{Higher Order Gravitational
  Deflection and Soft Bremsstrahlung in Planckian Energy Superstring
  Collisions},''
\href{http://dx.doi.org/10.1016/0550-3213(90)90375-N}{{\em Nucl. Phys.} {\bf
  B347} (1990)  550--580}.

\bibitem{Bellini:1992eb}
A.~Bellini, M.~Ademollo, and M.~Ciafaloni, ``{Superstring one loop and
  gravitino contributions to Planckian scattering},''
  \href{http://dx.doi.org/10.1016/0550-3213(93)90238-K}{{\em Nucl. Phys.} {\bf
  B393} (1993)  79--94},
\href{http://arxiv.org/abs/hep-th/9207113}{{\tt arXiv:hep-th/9207113
  [hep-th]}}.

\bibitem{DiVecchia:2019kta}
P.~Di~Vecchia, S.~G. Naculich, R.~Russo, G.~Veneziano, and C.~D. White, ``{A
  tale of two exponentiations in $ \mathcal{N} $ = 8 supergravity at subleading
  level},'' \href{http://dx.doi.org/10.1007/JHEP03(2020)173}{{\em JHEP} {\bf
  03} (2020)  173}, \href{http://arxiv.org/abs/1911.11716}{{\tt
  arXiv:1911.11716 [hep-th]}}.

\bibitem{Bern:2020gjj}
Z.~Bern, H.~Ita, J.~Parra-Martinez, and M.~S. Ruf, ``{Universality in the
  classical limit of massless gravitational scattering},''
  \href{http://dx.doi.org/10.1103/PhysRevLett.125.031601}{{\em Phys. Rev.
  Lett.} {\bf 125} (2020) no.~3, 031601},
  \href{http://arxiv.org/abs/2002.02459}{{\tt arXiv:2002.02459 [hep-th]}}.

\bibitem{Cheung:2018wkq}
C.~Cheung, I.~Z. Rothstein, and M.~P. Solon, ``{From Scattering Amplitudes to
  Classical Potentials in the Post-Minkowskian Expansion},''
  \href{http://dx.doi.org/10.1103/PhysRevLett.121.251101}{{\em Phys. Rev.
  Lett.} {\bf 121} (2018) no.~25, 251101},
\href{http://arxiv.org/abs/1808.02489}{{\tt arXiv:1808.02489 [hep-th]}}.

\bibitem{Bjerrum-Bohr:2018xdl}
N.~E.~J. Bjerrum-Bohr, P.~H. Damgaard, G.~Festuccia, L.~Plant{\'e}, and
  P.~Vanhove, ``{General Relativity from Scattering Amplitudes},''
  \href{http://dx.doi.org/10.1103/PhysRevLett.121.171601}{{\em Phys. Rev.
  Lett.} {\bf 121} (2018) no.~17, 171601},
\href{http://arxiv.org/abs/1806.04920}{{\tt arXiv:1806.04920 [hep-th]}}.

\bibitem{Bern:2019nnu}
Z.~Bern, C.~Cheung, R.~Roiban, C.-H. Shen, M.~P. Solon, and M.~Zeng,
  ``{Scattering Amplitudes and the Conservative Hamiltonian for Binary Systems
  at Third Post-Minkowskian Order},''
  \href{http://dx.doi.org/10.1103/PhysRevLett.122.201603}{{\em Phys. Rev.
  Lett.} {\bf 122} (2019) no.~20, 201603},
\href{http://arxiv.org/abs/1901.04424}{{\tt arXiv:1901.04424 [hep-th]}}.

\bibitem{KoemansCollado:2019ggb}
A.~Koemans~Collado, P.~Di~Vecchia, and R.~Russo, ``{Revisiting the second
  post-Minkowskian eikonal and the dynamics of binary black holes},''
  \href{http://dx.doi.org/10.1103/PhysRevD.100.066028}{{\em Phys. Rev. D} {\bf
  100} (2019) no.~6, 066028}, \href{http://arxiv.org/abs/1904.02667}{{\tt
  arXiv:1904.02667 [hep-th]}}.

\bibitem{Bern:2019crd}
Z.~Bern, C.~Cheung, R.~Roiban, C.-H. Shen, M.~P. Solon, and M.~Zeng, ``{Black
  Hole Binary Dynamics from the Double Copy and Effective Theory},''
\href{http://arxiv.org/abs/1908.01493}{{\tt arXiv:1908.01493 [hep-th]}}.

\bibitem{Bjerrum-Bohr:2019kec}
N.~E.~J. Bjerrum-Bohr, A.~Cristofoli, and P.~H. Damgaard, ``{Post-Minkowskian
  Scattering Angle in Einstein Gravity},''
\href{http://arxiv.org/abs/1910.09366}{{\tt arXiv:1910.09366 [hep-th]}}.

\bibitem{Kalin:2019rwq}
G.~K{\"a}lin and R.~A. Porto, ``{From Boundary Data to Bound States},''
  \href{http://dx.doi.org/10.1007/JHEP01(2020)072}{{\em JHEP} {\bf 01} (2020)
  072}, \href{http://arxiv.org/abs/1910.03008}{{\tt arXiv:1910.03008
  [hep-th]}}.

\bibitem{Kalin:2019inp}
G.~K{\"a}lin and R.~A. Porto, ``{From boundary data to bound states. Part II.
  Scattering angle to dynamical invariants (with twist)},''
  \href{http://dx.doi.org/10.1007/JHEP02(2020)120}{{\em JHEP} {\bf 02} (2020)
  120}, \href{http://arxiv.org/abs/1911.09130}{{\tt arXiv:1911.09130
  [hep-th]}}.

\bibitem{Cristofoli:2020uzm}
A.~Cristofoli, P.~H. Damgaard, P.~Di~Vecchia, and C.~Heissenberg,
  ``{Second-order Post-Minkowskian scattering in arbitrary dimensions},''
  \href{http://dx.doi.org/10.1007/JHEP07(2020)122}{{\em JHEP} {\bf 07} (2020)
  122}, \href{http://arxiv.org/abs/2003.10274}{{\tt arXiv:2003.10274
  [hep-th]}}.

\bibitem{Damour:2017zjx}
T.~Damour, ``{High-energy gravitational scattering and the general
relativistic
  two-body problem},''
\href{http://dx.doi.org/10.1103/PhysRevD.97.044038}{{\em
  Phys. Rev.} {\bf D97} (2018) no.~4, 044038},
\href{http://arxiv.org/abs/1710.10599}{{\tt arXiv:1710.10599 [gr-qc]}}.


\bibitem{Cheung:2020gyp}
C.~Cheung and M.~P. Solon, ``{Classical gravitational scattering at $
  \mathcal{O} $(G$^{3}$) from Feynman diagrams},''
  \href{http://dx.doi.org/10.1007/JHEP06(2020)144}{{\em JHEP} {\bf 06} (2020)
  144}, \href{http://arxiv.org/abs/2003.08351}{{\tt arXiv:2003.08351
  [hep-th]}}.

\bibitem{Kalin:2020fhe}
G.~K{\"a}lin, Z.~Liu, and R.~A. Porto, ``{Conservative Dynamics of Binary
  Systems to Third Post-Minkowskian Order from the Effective Field Theory
  Approach},'' \href{http://arxiv.org/abs/2007.04977}{{\tt arXiv:2007.04977
  [hep-th]}}.

\bibitem{Parra-Martinez:2020dzs}
J.~Parra-Martinez, M.~S. Ruf, and M.~Zeng, ``{Extremal black hole scattering at
  {$O(G^3)$}: graviton dominance, eikonal exponentiation, and differential
  equations},'' \href{http://arxiv.org/abs/2005.04236}{{\tt arXiv:2005.04236
  [hep-th]}}.

\bibitem{Weinberg:1965nx}
S.~Weinberg, ``{Infrared photons and gravitons},''
\href{http://dx.doi.org/10.1103/PhysRev.140.B516}{{\em Phys. Rev.} {\bf 140}
  (1965)  B516--B524}.

\bibitem{Damour:2019lcq}
T.~Damour, ``{Classical and quantum scattering in post-Minkowskian gravity},''
  \href{http://dx.doi.org/10.1103/PhysRevD.102.024060}{{\em Phys. Rev. D} {\bf
  102} (2020) no.~2, 024060}, \href{http://arxiv.org/abs/1912.02139}{{\tt
  arXiv:1912.02139 [gr-qc]}}.
  
  \bibitem{Blumlein:2020znm}
J.~Bl{\"{u}}mlein, A.~Maier, P.~Marquard and G.~Sch{\"{a}}fer,
``Testing binary dynamics in gravity at the sixth post-Newtonian level,''
Phys. Lett. B \textbf{807}, 135496 (2020)
[arXiv:2003.07145 [gr-qc]].

\bibitem{Bini:2020wpo}
D.~Bini, T.~Damour and A.~Geralico,
``Binary dynamics at the fifth and fifth-and-a-half post-Newtonian orders,''
Phys. Rev. D \textbf{102}, no.2, 024062 (2020)
[arXiv:2003.11891 [gr-qc]].


\bibitem{Amati:2007ak}
D.~Amati, M.~Ciafaloni, and G.~Veneziano, ``{Towards an S-matrix Description of
  Gravitational Collapse},''
  \href{http://dx.doi.org/10.1088/1126-6708/2008/02/049}{{\em JHEP} {\bf 02}
  (2008)  049},
\href{http://arxiv.org/abs/0712.1209}{{\tt arXiv:0712.1209 [hep-th]}}.

\bibitem{Ademollo:1990sd}
M.~Ademollo, A.~Bellini, and M.~Ciafaloni, ``{Superstring Regge amplitudes and
  graviton radiation at planckian energies},''
\href{http://dx.doi.org/10.1016/0550-3213(90)90626-O}{{\em Nucl.Phys.} {\bf
  B338} (1990)  114--142}.

\bibitem{toap}
P.~Di~Vecchia, C.~Heissenberg, R.~Russo, and G.~Veneziano, to appear.

\bibitem{Bronzan:1974jh}
J.~Bronzan, G.~L. Kane, and U.~P. Sukhatme, ``{Obtaining Real Parts of
  Scattering Amplitudes Directly from Cross-Section Data Using Derivative
  Analyticity Relations},''
  \href{http://dx.doi.org/10.1016/0370-2693(74)90432-8}{{\em Phys. Lett. B}
  {\bf 49} (1974)  272--276}.

\bibitem{Kang:1974gt}
K.~Kang and B.~Nicolescu, ``{Models for Hadron - Hadron Scattering at
  High-Energies and Rising Total Cross-Sections},''
  \href{http://dx.doi.org/10.1103/PhysRevD.11.2461}{{\em Phys. Rev. D} {\bf 11}
  (1975)  2461}.

\bibitem{Fagundes:2017iwb}
D.~Fagundes, M.~Menon, and P.~Silva, ``{Leading components in forward elastic
  hadron scattering: Derivative dispersion relations and asymptotic
  uniqueness},'' \href{http://dx.doi.org/10.1142/S0217751X17501846}{{\em Int.
  J. Mod. Phys. A} {\bf 32} (2017) no.~32, 1750184},
  \href{http://arxiv.org/abs/1705.01504}{{\tt arXiv:1705.01504 [hep-ph]}}.

\bibitem{Kabat:1992tb}
D.~N. Kabat and M.~Ortiz, ``{Eikonal quantum gravity and Planckian
  scattering},'' \href{http://dx.doi.org/10.1016/0550-3213(92)90627-N}{{\em
  Nucl. Phys.} {\bf B388} (1992)  570--592},
\href{http://arxiv.org/abs/hep-th/9203082}{{\tt arXiv:hep-th/9203082
  [hep-th]}}.

\bibitem{Henn:2019rgj}
J.~M. Henn and B.~Mistlberger, ``{Four-graviton scattering to three loops in $
  \mathcal{N}=8 $ supergravity},''
  \href{http://dx.doi.org/10.1007/JHEP05(2019)023}{{\em JHEP} {\bf 05} (2019)
  023},
\href{http://arxiv.org/abs/1902.07221}{{\tt arXiv:1902.07221 [hep-th]}}.

\bibitem{DiVecchia:2019myk}
P.~Di~Vecchia, A.~Luna, S.~G. Naculich, R.~Russo, G.~Veneziano, and C.~D.
  White, ``{A tale of two exponentiations in ${\cal N}=8$ supergravity},''
  \href{http://dx.doi.org/10.1016/j.physletb.2019.134927}{{\em Phys. Lett.}
  {\bf B798} (2019)  134927},
\href{http://arxiv.org/abs/1908.05603}{{\tt arXiv:1908.05603 [hep-th]}}.

\bibitem{BjerrumBohr:2010zs}
N.~Bjerrum-Bohr, P.~H. Damgaard, T.~Sondergaard, and P.~Vanhove, ``{Monodromy
  and Jacobi-like Relations for Color-Ordered Amplitudes},''
  \href{http://dx.doi.org/10.1007/JHEP06(2010)003}{{\em JHEP} {\bf 06} (2010)
  003}, \href{http://arxiv.org/abs/1003.2403}{{\tt arXiv:1003.2403 [hep-th]}}.

\bibitem{Caron-Huot:2018ape}
S.~Caron-Huot and Z.~Zahraee, ``{Integrability of Black Hole Orbits in Maximal
  Supergravity},'' \href{http://dx.doi.org/10.1007/JHEP07(2019)179}{{\em JHEP}
  {\bf 07} (2019)  179},
\href{http://arxiv.org/abs/1810.04694}{{\tt arXiv:1810.04694 [hep-th]}}.

\bibitem{Bianchi:2016yiq}
M.~S. Bianchi and M.~Leoni, ``{A $QQ\to QQ$ planar double box in canonical
  form},'' \href{http://dx.doi.org/10.1016/j.physletb.2017.12.030}{{\em Phys.
  Lett. B} {\bf 777} (2018)  394--398},
  \href{http://arxiv.org/abs/1612.05609}{{\tt arXiv:1612.05609 [hep-ph]}}.

\bibitem{Henn:2013woa}
J.~M. Henn and V.~A. Smirnov, ``{Analytic results for two-loop master integrals
  for Bhabha scattering I},''
  \href{http://dx.doi.org/10.1007/JHEP11(2013)041}{{\em JHEP} {\bf 11} (2013)
  041}, \href{http://arxiv.org/abs/1307.4083}{{\tt arXiv:1307.4083 [hep-th]}}.

\bibitem{Smirnov:2001cm}
V.~A. Smirnov, ``{Analytical result for dimensionally regularized massive
  on-shell planar double box},''
  \href{http://dx.doi.org/10.1016/S0370-2693(01)01382-X}{{\em Phys. Lett. B}
  {\bf 524} (2002)  129--136}, \href{http://arxiv.org/abs/hep-ph/0111160}{{\tt
  arXiv:hep-ph/0111160}}.

\bibitem{Heinrich:2004iq}
G.~Heinrich and V.~A. Smirnov, ``{Analytical evaluation of dimensionally
  regularized massive on-shell double boxes},''
  \href{http://dx.doi.org/10.1016/j.physletb.2004.07.058}{{\em Phys. Lett. B}
  {\bf 598} (2004)  55--66}, \href{http://arxiv.org/abs/hep-ph/0406053}{{\tt
  arXiv:hep-ph/0406053}}.


\end{thebibliography}
\end{document}